\documentclass[letterpaper]{IEEEtran}
\usepackage{graphicx,times, amsmath, amsfonts,comment}
\usepackage{amssymb,epstopdf}
\usepackage[noend]{algorithmic}
\usepackage{algorithm}

\newcommand{\beq}{\begin{equation}}
\newcommand{\eeq}{\end{equation}}

\newcommand{\ud}{\mathrm{d}}

\newcommand {\Ebb}{\mathbb{E}}

\begin{document}

\title{Saddle Point Approximation for Outage Probability Using Cumulant Generating Functions}
\author{\IEEEauthorblockN{Sudarshan Guruacharya, Hina Tabassum, and Ekram Hossain \\} 
\IEEEauthorblockA{Department of Electrical and Computer Engineering, University of Manitoba, Canada \\
Emails: \{guruachs, hina.tabassum, ekram.hossain\}@umanitoba.ca}}
\maketitle

\begin{abstract}
This letter proposes the use of  saddle point approximation (SPA) to evaluate the  outage probability of wireless cellular networks. Unlike traditional numerical integration-based approaches, the SPA approach relies on cumulant generating functions (CGFs)  and eliminates the need for explicit numerical integration.  The approach is generic and can be applied to a wide variety of distributions, given that their CGFs exist. We illustrate the usefulness of SPA on channel fading distributions such as Nakagami-$m$, Nakagami-$q$ (Hoyt), and Rician distributions. Numerical results validate the accuracy of the proposed SPA approach.
\end{abstract}


\section{Introduction}
Given the instantaneous signal power $p_0$ and interfering signal powers $p_k$ from $k = 1,\ldots, L$ interferers at a wireless receiver, the signal-to-interference ratio (SIR) at the receiver is defined as $\mathrm{SIR} = p_0/I$, where $I$ is the aggregate interference at the receiver from $L$ interferers, i.e., $I = \sum_{k=1}^L p_k$. By definition, the SIR outage occurs when $qI>p_0$, where $q$ is the desired SIR threshold. Introducing a new random variable $\gamma = qI - p_0$, the SIR outage probability can be given as
\beq
P_{\mathrm{out}} = \mathrm{Pr}(\gamma > 0) = Q_\gamma(0),
\label{eqn:outage-def}
\eeq 
where $Q_\gamma$ is the complementary cumulative distribution function (CCDF) of $\gamma$.\footnote{Note that signal-to-interference-plus-noise ratio (SINR) outage can also be expressed in a similar manner. Since $\mathrm{SINR} = p_0/(I + N_0)$,  where $N_0$ is the noise power, the SINR outage can be  given in terms of $\gamma$ as $\mathrm{Pr}(\gamma > -qN_0)$.}

Given that $\gamma$ is a linear combination of independent random variables $p_k$, we can use the product of the moment generating functions (MGF) of the random variables to obtain the MGF of $\gamma$. The SIR outage probability in (\ref{eqn:outage-def}) can then be evaluated using Gil-Pelaez inversion theorem as~\cite{Gil-Pelaez1951}:
\beq
Q_{\gamma}(x) = \frac{1}{2} + \frac{1}{\pi} \int_0^\infty \frac{\mathrm{Im}\{M_\gamma(jt) e^{-jtx}\}}{t} \ud t,
\label{eqn:gil-pelaez-outage}
\eeq
where $M_\gamma$ is the MGF of $\gamma$ and $\mathrm{Im\{z\}}$ is the imaginary component of complex variable $z$. 
This MGF approach using an intermediate variable $\gamma$ is both theoretically elegant over direct evaluation of the CDF of SIR as well as numerically advantageous over computationally intensive Monte-Carlo methods.

The MGF approach was first described in \cite{Zhang1995} and was applied to Nakagami-$m$ fading channels. Since then the integral in (\ref{eqn:gil-pelaez-outage})  has been investigated in several studies. For some distributions, it is possible to analytically evaluate  (\ref{eqn:gil-pelaez-outage}) using residue calculus of complex analysis. Such an attempt was made in  \cite{Zhang1995} and \cite{Sharma2007} for Nakagami-$m$ fading channels. For arbitrary distributions, the integral can be evaluated numerically. In \cite{Zhang1996},  (\ref{eqn:gil-pelaez-outage}) was transformed into polar coordinates after which the limits of the integral become finite. This finite form is much more amenable to numerical integration techniques, and it became the basis for subsequent developments in the numerical evaluation of the outage probability. In \cite{Zhang1996},  the use of Gaussian quadrature was suggested. This numerical approach was generalized to other distributions in \cite{Annamalai2001}, which championed the use of Gauss-Chebyshev quadrature. In \cite{Young-chai2000}, the authors suggested the use of trapezoidal rule and Euler summation. A numerical contour integration method, that approximates the steepest descent path, was proposed in \cite{Senaratne2009}; but the authors missed the Lugannani-Rice formula already well established in the literature.

This letter supplements the aforementioned numerical approaches by proposing the use of saddle point approximation (SPA) for outage evaluations. The SPA approach relies on cumulant generating functions (CGFs) and eliminates the need for explicit numerical integration. The integration problem is instead substituted by a minimization problem. The method is generic and can handle a wide variety of distributions.  To the best of our knowledge, the use of SPA has not been explored for outage computations. The possibility to use this approach to compute other metrics of interest such as bit error rate and ergodic capacity\footnote{The ergodic capacity is related to SINR outage by $\Ebb[C] = \int_0^\infty (1- P_{out}) \;\ud q$, where $C = \log_2(1+{\rm SINR})$.} is also obvious. While the technique is generally valid, in this letter, we focus on well-known  channel fading distributons such as Nakagami-m, Nakagami-q (or Hoyt), and Rice distributions. These distributions can model the empirical fast fading measurements quite well and reduce to other simple fading distributions such as Rayleigh as their special cases.


\section{Saddle Point Approximation}
SPA (also known as the method of steepest descent) is a powerful method of obtaining asymptotic approximations to Laplace type integrals of the form $I(z) = \int_a^b t^{\alpha - 1} e^{z f(t)} g(t) \;\ud t$ as $z \rightarrow \infty$. The method was introduced in \cite{Daniels1954} to approximate the PDF of sum of independent and identically distributed (IID) random variables. Based on Gil-Pelaez's inversion formula, the Lugannani-Rice formula was derived to approximate the CDF of sum of IID random variables in \cite{Lugannani-Rice1980}. As the number of variables increase, the better is the approximation. However, the method satisfactorily approximates the CDF of a single random variable as well, which is the reason for its great appeal. An easier derivation of Lugannani-Rice formula, which is more statistically flavored, was given in \cite{Daniels1987}. An introduction to these techniques can be found in \cite{Goutis1999} and \cite[Ch. 32]{Gentle2012}, while a definitive exposition can be found in \cite{Butler2007}. 

For an arbitrary random variable $X$, the SPA approach requires its MGF as well as CGF to exist. The MGF is given as $M_X(t) = \Ebb[e^{tx}]$. Since the integrals need not always converge absolutely, $M_X(t)$ may not exist. $M_X(t)$ exists if it is finite around the neighborhood of zero, i.e., if there exists an $h > 0$ such that $\forall t \in (-h,h)$, $M_X(t) < \infty$. If $M_X(t)$ exists, the largest open interval $U$ around zero such that $M_X(t) < \infty$ for $t\in U$ is referred to as the the convergence strip of the MGF of $X$. Also, if $M_X(t)$ exists, then all positive moments of $X$ exist as well. 

Using SPA approach, the evaluation of CCDF of an arbitrary random variable $X$ requires the following steps:
\begin{enumerate}
\item Deriving the CGF which is defined as the natural logarithm of MGF, $K_X(t) = \log M_X(t)$. If CGF exists, then it is always a convex function that passes through the origin. Similar to MGF, the CGF allows us to obtain the $n$-th cumulant of $X$ by evaluating its $n$-th derivative at zero, $\kappa_n = K^{(n)}(0)$. The first two cumulants are the mean and the variance.

\item Given the CGF $K_X$, finding the saddle point $\hat{t}(x)$ which solves the following saddle point equation:
\beq 
K_X'(\hat{t}) = x. 
\label{eqn:saddle-pt-eq}
\eeq

 \item The CCDF of $X$ can then be approximated by the first three terms of the Lugannani-Rice (LR) formula as 
\beq 
Q_X(x) = 1 - \Phi(w) + \phi(w) \Big(\frac{1}{u} - \frac{1}{w} \Big), 
\label{eqn:LR-formula}
\eeq
where $Q_X(x)$ is the CCDF of $X$, $\Phi$ and $\phi$ are CDF and PDF of standard normal distribution, respectively, and 
 \begin{align*}
 w &= \mathrm{sign}(\hat{t}) \sqrt{2(x\hat{t} - K_X(\hat{t}))}, \\
 u &= \hat{t} \sqrt{K_X''(\hat{t})}, 
 \end{align*}
 where $\mathrm{sign}(x)$ is 1 if $x\geq0$ and 0 if $x<0$. If $x=\Ebb[X]$, then $\hat{t} = 0$ is the saddle point for $\Ebb[X]$, since by definition of CGF $K_X'(0) = \Ebb[X]$. Hence, at the mean, $K_X(0) = 0$, which leads to $\hat{w} = 0$, making (\ref{eqn:LR-formula}) inapplicable. When $x=\Ebb[X]$, the corresponding value of CCDF is 
 \beq
 Q_X(x) = \frac{1}{2} - \frac{K_X'''(0)}{6\sqrt{2\pi}K_X''(0)^{3/2}}. 
 \eeq
 However, in practice it is easier to linearly interpolate the value of $Q_X$ based on (\ref{eqn:LR-formula}) to $\Ebb[X] \pm \epsilon$, where $\epsilon$ is a sufficiently small value. The relative algebraic simplicity of (\ref{eqn:LR-formula}) has made it popular in many applications.
\end{enumerate}


\section{SPA for Evaluation of SIR Outage Probability}
For the calculation of SIR outage probability using SPA, the MGF and CGF of $\gamma$ is given by $M_\gamma(t) = M_I(qt)M_{P_0}(-t)$ and $K_\gamma(t) = \log M_I(qt) + \log M_{P_0}(-t)$, respectively. For independent interferers, $M_I(t) = \prod_{k=1}^L M_{P_k}(t)$ and $K_I(t) = \sum_{k=1}^L \log M_{P_k}(t)$. Consequently, we can obtain the SPA of $Q_\gamma(0)$ by first solving the saddle point equation (\ref{eqn:saddle-pt-eq}) and then substituting the value of $\hat{t}$ in the LR formula (\ref{eqn:LR-formula}). For the random variable $\gamma$, putting $x = 0$ in (\ref{eqn:saddle-pt-eq})  gives 
 \beq 
 K_\gamma'(t) = 0. 
 \label{eqn:saddle-pt-eq-outage}
 \eeq 
 Similarly, the expression for $w$ simplifies to $ w = \mathrm{sign}(\hat{t}) \sqrt{-2 K_\gamma(\hat{t})}$. Thus, a valid $\hat{t}(0)$ should give $K_\gamma(\hat{t}) < 0$ and $K_\gamma''(\hat{t}) >0$. As mentioned previously, the LR formula in (\ref{eqn:LR-formula}) will breakdown for $x = 0$ when $\Ebb[\gamma]=x$, i.e., when $q \sum_{k=1}^L \bar{p}_k = \bar{p}_0$, where $\bar{p}_k$ is the average signal power. For identical interferers, when $\bar{p}_k = \bar{p}$ for $k=1,\ldots, L$, we can express the threshold at which this breakdown occurs in dB as $q_{\mathrm{dB}} = \bar{p}_{0,\mathrm{dBm}} - \bar{p}_{\mathrm{dBm}} - L$. In such a case, we linearly interpolate the value of $Q_\gamma(0)$.\footnote{The SINR outage can likewise be obtained by solving $K'_\gamma(t) = -qN_0$.}
 
Depending on the complexity of MGF, it may not be possible to obtain an explicit analytical expression for $\hat{t}(0)$ that solves the saddle point equation (\ref{eqn:saddle-pt-eq-outage}). Thus, we need to resort to numerical root finding techniques. While any root finding method suffices, for our purpose, we will approximate the saddle point using the Newton-Raphson method. The $n$-th iterate of Newton-Raphson method is given by 
 \beq 
 \hat{t}_{n+1} = \hat{t}_n - \frac{K_\gamma'(\hat{t}_n)}{K_\gamma''(\hat{t}_n)},
 \label{eqn:newton-ralphson}
 \eeq
where we can initialize $\hat{t}_0 = 0$. For this initial value, we have by definition $K_\gamma(0) = 0$, $K'_\gamma(0) = \Ebb[\gamma]$, and $K''_\gamma(0) = \mathrm{Var}[\gamma]$. We can iterate until the approximation error is below a suitable tolerance. Since the CGF $K_\gamma$ is by definition a convex function, when $x=0$ the Netwon-Raphson method essentially finds the unique global minima of $K_\gamma$. 

Since CGF of the sum of independent random variables is given by the sum of CGFs of all the random variables, the first and second derivatives of $\gamma$ required for (\ref{eqn:newton-ralphson}) can be computed term wise for each variable $p_k$ and then added up. First and second derivative of $K_{P_k}$ can be given in terms of $M_{P_k}$ as $K'_{P_k}(t) = \frac{M'_{P_k}(t)}{M_{P_k}(t)}$ and $K''_{P_k}(t) = \frac{M''_{P_k}(t) M_{P_k}(t) - M'_{P_k}(t)^2}{M_{P_k}(t)^2}$. Thus the required derivatives of $K_\gamma$ are 
\begin{align}
K'_\gamma(t) &= \sum_{k=1}^L K'_{P_k}(qt) + K'_{P_0}(-t) \label{eqn:1st-der-K} \\ 
K''_\gamma(t) &= \sum_{k=1}^L K''_{P_k}(qt) + K''_{P_0}(-t). \label{eqn:2nd-der-K}
\end{align}
Equation (\ref{eqn:saddle-pt-eq-outage}) can then be solved to obtain $\hat{t}$, after which it can be substituted in (\ref{eqn:LR-formula}) to evaluate the SIR outage probability.

In the rest of this section, we will describe the SPA approach when all the channels are independent and follows Nakagami-$m$, Nakagami-$q$ (or Hoyt), and Rice distributions. For simplicity, we will also assume that the interferers are identically distributed, although this is not a necessary assumption and the SPA approach is valid of non-identical interferers as well.


\subsection{Nakagami-$m$ Channel}
For Nakagami-$m$ channel, the PDF of signal power $p_k$ follows Gamma distribution given by 
 \[
 f_{p_k}(x) = \frac{\lambda_k^{m_k}}{\Gamma(m_k)} x^{m_k-1} \exp(-\lambda_k x), 
 \]
where $x \geq 0$. The $m_k \in [0.5, \infty)$ is the fading parameter and $\lambda_k$ is defined as $\lambda_k =m_k/\bar{p}_k$. Rayleigh fading is obtained when $m_k=1$. For the case with identical interferers, let $\lambda_k = \lambda$ and $m_k = m$ for $k = 1, \ldots, L$. 

For the Gamma distributed $p_k$'s, the MGF of $\gamma$ is given by 
\begin{equation*}
M_\gamma(t) = \left( 1 - \frac{qt}{\lambda} \right)^{- Lm} \left( 1 + \frac{t}{\lambda_0} \right)^{-m_0},
\end{equation*}
such that $|\frac{t}{\lambda_0}| < 1$ and $|\frac{qt}{\lambda}| < 1$. Thus, the convergence strip of $M_\gamma$ is given by $|t|< \min (\lambda_0,\frac{\lambda}{q})$. Taking the logarithm of $M_\gamma(t)$, we have the CGF and its derivative as 
 \begin{align*} 
K_\gamma(t) &= - L m \log \left(1 - \frac{qt}{\lambda} \right) - m_0 \log \left(1 + \frac{t}{\lambda_0} \right), \\
 K_\gamma'(t) &= - \frac{Lm}{t - \lambda/q} - \frac{m_0}{t + \lambda_0}.
 \end{align*}

For Nakagami-$m$ fading with identical interferers, we can explicitly solve the saddle point equation (\ref{eqn:saddle-pt-eq-outage}) to obtain 
\beq
\hat{t}(0) = \frac{m_0 \lambda/q - mL\lambda_0}{mL + m_0}.
\label{eqn:SPA-iid-nakagami-gamma-zero}
\eeq
Substituting the value of $\hat{t}$ given by (\ref{eqn:SPA-iid-nakagami-gamma-zero}) in (\ref{eqn:LR-formula}) gives us the required SIR outage. Thus, the SPA for Nakagami-$m$ fading with identical interferers can be given in closed form.


\subsection{Rician Channel}
For Rician channel, the PDF of signal power $p_k$ is given by
\[ 
f_{p_k}(x) = \frac{1 + r_k}{\bar{p}_k} \exp\left[ - r_k - \frac{(1+r_k)x}{\bar{p}_k}\right] I_0 \left[ 2 \sqrt{\frac{r_k (r_k + 1)x}{\bar{p}_k}}\right],
\]
where $x \ge 0$, $r_k \geq 0$ is the Rice parameter, and $I_0(\cdot)$ is the modified Bessel function of the first kind with order zero. Rayleigh fading is obtained when $r_k = 0$. The corresponding MGF of $p_k$ is 
\[ 
M_{P_k}(t) = \frac{1 + r_k}{1 + r_k - t\bar{p}_k} \exp \left( \frac{r_k \bar{p}_k t}{1 + r_k - t \bar{p}_k} \right), 
\]
such that $|\frac{t\bar{p}_k}{1+r_k}|<1$. Taking its logarithm, the CGF and its derivatives are given by
\begin{align*}
K_{P_k}(t) &= \log(1 + r_k) - \log(1 + r_k - t \bar{p}_k) + \frac{r_k \bar{p}_k t}{1 + r_k - \bar{p}_k t}, \\
K'_{P_k}(t) &= \bar{p}_k \frac{(1+r_k)^2 - \bar{p}_k t}{(1 + r_k - \bar{p}_k t)^2}, \\
K''_{P_k}(t) &= \bar{p}_k^2 \frac{2 r_k^2 + 3 r_k - \bar{p}_k t + 1}{(1 + r_k - \bar{p}_k t)^3}.
\end{align*}

For the case with identical interferers, let $r_k = r$ and $\bar{p}_k = \bar{p}$ for $k = 1, \ldots, L$. The convergence strip is given by $|t| < \min(\frac{1+r}{\bar{p} q}, \frac{1+r_0}{\bar{p}_0})$. The saddle point equation (\ref{eqn:saddle-pt-eq-outage}) then becomes a cubic polynomial. Although it is possible to solve a cubic polynomial analytically, in our context we use the Newton-Raphson method in (\ref{eqn:newton-ralphson}) to obtain the saddle point. The first and second derivatives of $K_\gamma$ can be computed as in (\ref{eqn:1st-der-K}) and (\ref{eqn:2nd-der-K}), respectively. Finally, substituting the  obtained value for $\hat{t}$ in the LR formula in (\ref{eqn:LR-formula}) gives us the required SIR outage probability.


\subsection{Nakagami-$q$ (or Hoyt) Channel}
For Nakagami-$q$ channel, the PDF of signal power $p_k$ is given by 
\[
f_{p_k}(x) = \frac{1}{\bar{p}_k \sqrt{1-b_k^2}}\exp\left[ \frac{-x}{(1-b_k^2)\bar{p}_k} \right] I_0 \left[ \frac{b_k x}{(1-b_k^2)\bar{p}_k} \right],
\]
where $x\geq0$ and $-1 \leq b_k = \frac{1-m_k^2}{1+m_k^2} \leq 1$. The $m_k \in [0, \infty)$ is the fading parameter. Rayleigh fading occurs when $b_k = 0$. The MGF of $p_k$ is given by
\[
 M_{P_k}(t) =\frac{1}{\sqrt{(1 - t \bar{p}_k (1 + b_k))(1 - t \bar{p}_k (1 - b_k) )}},
\]
such that $|t \bar{p}_k (1 + b_k)| < 1$ and $|t \bar{p}_k (1 - b_k)| < 1$. Taking its logarithm, the CGF and its derivatives are
\begin{align*}
K_{P_k}(t) &= - \frac{1}{2}\log(1 - t \bar{p}_k (1-b_k)) -\frac{1}{2}\log(1 - t \bar{p}_k (1+b_k)), \\
K'_{P_k}(t) &= \frac{(1-b_k)\bar{p}_k}{2(1 - (1-b_k)\bar{p}_k t)} + \frac{(1+b_k)\bar{p}_k }{2(1 - (1+b_k)\bar{p}_k t )}, \\
K''_{P_k}(t) &= \frac{(1-b_k)^2 \bar{p}_k^2}{2(1 - (1-b_k)\bar{p}_k t)^2} + \frac{(1+b_k)^2 \bar{p}_k^2}{2(1 - (1+b_k)\bar{p}_k t)^2}.
\end{align*}

For the case with identical interferers, let $b_k = b$ and $\bar{p}_k = \bar{p}$ for $k = 1, \ldots, L$. The convergence strip is given by $|t| < \min(t_I, t_{P_0})$, where $t_I = \min(\frac{1}{\bar{p} q (1+b)}, \frac{1}{\bar{p} q (1-b)})$ and $t_{P_0} = \min(\frac{1}{\bar{p}_0 (1+b_0)}, \frac{1}{\bar{p}_0 (1-b_0)})$. The saddle point equation (\ref{eqn:saddle-pt-eq-outage}) again becomes a cubic polynomial. Therefore, the procedure mentioned for Rician fading channels can again be applied for Nakagami-$q$ channels.


\section{Numerical Illustrations}
\begin{figure}[h]
\begin{center}
	\includegraphics[width=3in]{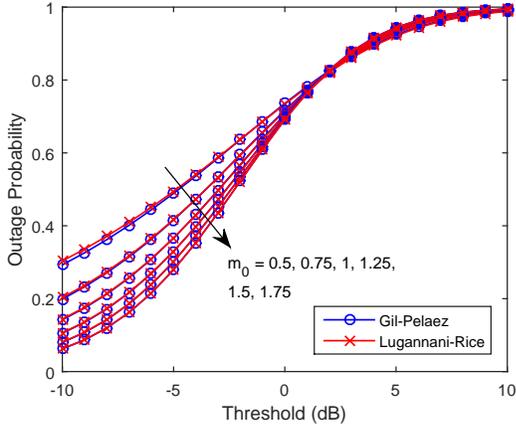}
	\caption{Outage probability for Nakagami-$m$ channel with identical interferers. }
	\label{fig:iid-nakagami-m}
 \end{center}
\end{figure}

\begin{figure}[h]
\begin{center}
	\includegraphics[width=3in]{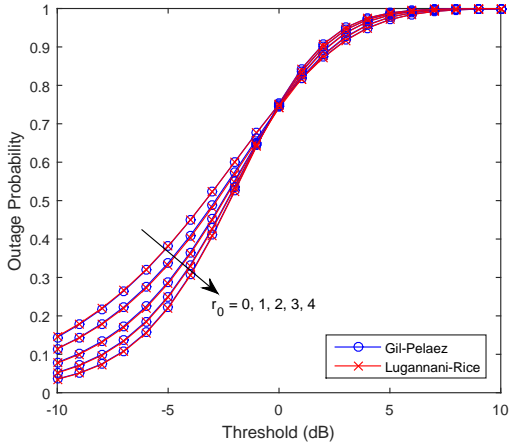}
	\caption{Outage probability for Rician channel with identical interferers. }
	\label{fig:iid-rice}
 \end{center}
\end{figure}

\begin{figure}[h]
\begin{center}
	\includegraphics[width=3in]{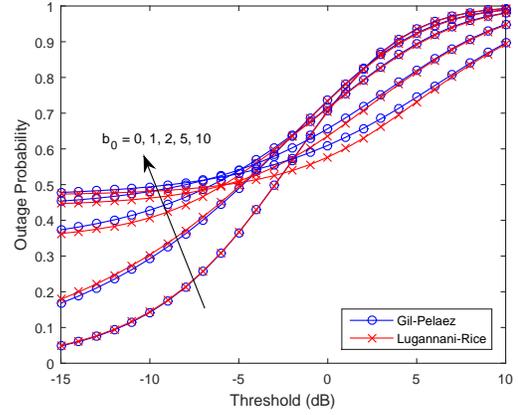}
	\caption{Outage probability for Nakagami-$q$ channel with identical interferers.}
	\label{fig:iid-nakagami-q}
 \end{center}
\end{figure}

In the following, we plot the SIR outage as a function of SIR threshold $q$. We set $\bar{p}_0 = 5$ dBm, $\bar{p} = 0$ dBm, and $L = 5$ for each plot; and the interferers are identically distributed.  The tolerance for the Newton-Raphson method is set to $10^{-8}$, which is typically achieved within 6 to 10 iterations. In Fig. \ref{fig:iid-nakagami-m}, we plot the results for Nakagami-$m$ channels.
We set $m = 0.5$ and $m_0$ is varied as $m_0 = 0.5, 0.75, 1, 1.25, 1.5, 1.75$. In Fig. \ref{fig:iid-rice}, we plot the results for Rician channels where we set the Rice parameters to $r = 0.5$ while $r_0$ is varied as $r_0 = 0, 1, 2, 3, 4$. In Fig. \ref{fig:iid-nakagami-q}, we plot the results for Nakagami-$q$ channels where we set the fading parameter to $b=1$ and $b_0$ is varied as $b_0 = 0, 1, 2, 5, 10$. We see that the results from SPA agree with the results obtained from numerical integration of Gil-Pelaez formula. For the case of Nakagami-$q$ channels, we see that the error is large around the transition point at lower threshold values. This error becomes worse as the value of $b$ is increased. 

\begin{figure}[h]
\begin{center}
	\includegraphics[width=3in]{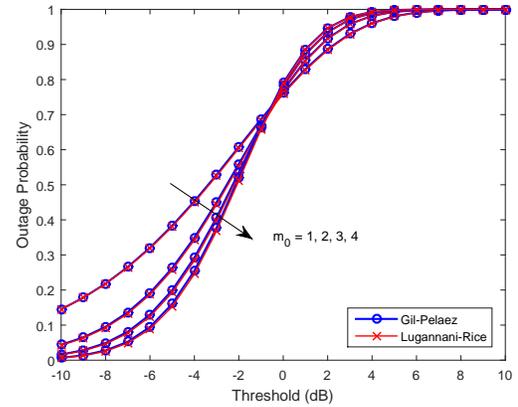}
	\caption{Outage probability for  Nakagami-$m$ channel with independent and non-identical interferers.}
	\label{fig:non-iid-nakagami-m}
 \end{center}
\end{figure}

To demonstrate the general validity of the method, we include Fig. \ref{fig:non-iid-nakagami-m} where we plot the outage versus threshold for Nakagami-$m$ channels when the interferers are non-identical. The number of interferers is $L = 4$. We set $\bar{p}_0 = 5$ dBm and $\bar{p}_k = 0$ dBm for all $k = 1,\ldots, L$. The fading parameters of the four interferers are assumed to be $m = [3.7, 3.5, 4.1, 1.7, 2.1]$ while $m_0$ is varied as $m_0 = 1, 2, 3, 4$. We see that the results obtained by numerically integrating Gil-Pelaez formula agree with those obtained from SPA.

\section{Conclusion}
We have discussed and demonstrated the utility of saddle point approximation for calculation of outage probability via Luganani-Rice formula. The Nakagami-$m$, Rice, and Nakagami-$q$ channels have been studied in detail.

\bibliographystyle{IEEE}

\end{document}